%% file: main.tex
\documentclass[conference]{IEEEtran}

\usepackage{cite}
\usepackage[table]{xcolor}
\usepackage{amsmath,amssymb,amsfonts}
\usepackage{algorithmic}
\usepackage{graphicx}
\usepackage{textcomp}
\usepackage{booktabs}
\usepackage{amsmath, balance}
\usepackage{scalerel}
\usepackage{tikz}
\usetikzlibrary{svg.path}
\usepackage{arydshln}
\usepackage{caption}
\usepackage{academicons}
\usepackage{array}  
\usepackage{lipsum}  
\usepackage{adjustbox}
\usepackage{mdframed}
\usepackage{multirow}

\definecolor{Gr}{rgb}{0.0, 0.5, 0.0}
\definecolor{light-gray}{gray}{0.95}
\newcommand{\code}[1]{\colorbox{light-gray}{\texttt{#1}}}

\newcommand{\specialcell}[2][c]{%
  \begin{tabular}[#1]{@{}c@{}}#2\end{tabular}}

\definecolor{purple}{HTML}{AF72B0}

\def\BibTeX{{\rm B\kern-.05em{\sc i\kern-.025em b}\kern-.08em
    T\kern-.1667em\lower.7ex\hbox{E}\kern-.125emX}}

\begin{document}

\title{AdvFusion: Adapter-based Knowledge Transfer for Code
Summarization on Code Language Models}






\makeatletter
\newcommand{\newlineauthors}{%
  \end{@IEEEauthorhalign}\hfill\mbox{}\par
  \mbox{}\hfill\begin{@IEEEauthorhalign}
}
\makeatother

\author{\IEEEauthorblockN{Iman Saberi}
\IEEEauthorblockA{\textit{The University of British Columbia, Okanagan} \\
Kelowna, Canada \\
iman.saberi@ubc.ca}
\and
\IEEEauthorblockN{Amirreza Esmaeili}
\IEEEauthorblockA{\textit{The University of British Columbia, Okanagan} \\
Kelowna, Canada \\
a.esmaeili@ubc.ca}

\newlineauthors
\IEEEauthorblockN{Fatemeh Fard}
\IEEEauthorblockA{\textit{The University of British Columbia, Okanagan} \\
Kelowna, Canada \\
fatemeh.fard@ubc.ca}

\and
\IEEEauthorblockN{Fuxiang Chen}
\IEEEauthorblockA{\textit{University of Leicester} \\
Leicester, United Kingdom \\
fuxiang.chen@leicester.ac.uk}
}

\maketitle

\input{Sections/abstract}


\input{Sections/introduction-FHF}

\input{Sections/background}

\input{Sections/Motivation}

\input{Sections/AdversarialFusion}

\input{Sections/ResearchQuestions}

\input{Sections/Results}

\input{Sections/Discussion.tex}
\input{Sections/RelatedWork.tex}
\input{Sections/Threats.tex}
\input{Sections/Conclusion.tex}

\section*{Acknowledgement}
This research is supported by a grant from the Natural Sciences and Engineering Research Council of Canada RGPIN-2019-05175.

\balance
\bibliographystyle{IEEEtran}
\bibliography{References}

\end{document}

%% file: Sections/abstract.tex
\begin{abstract}

Programming languages can benefit from one another by utilizing a pre-trained model for software engineering tasks such as code summarization and method name prediction. While full fine-tuning of Code Language Models (Code-LMs) has been explored for multilingual knowledge transfer, research on Parameter Efficient Fine-Tuning (PEFT) for this purpose is limited. AdapterFusion, a PEFT architecture, aims to enhance task performance by leveraging information from multiple languages but primarily focuses on the target language.

To address this, we propose \textbf{AdvFusion}, a novel PEFT-based approach that effectively learns from other languages before adapting to the target task. Evaluated on code summarization and method name prediction, AdvFusion outperforms AdapterFusion by up to 1.7 points and surpasses LoRA with gains of 1.99, 1.26, and 2.16 for Ruby, JavaScript, and Go, respectively. We open-source our scripts for replication purposes\footnote{https://github.com/ist1373/AdvFusion}.

\end{abstract}

%% file: Sections/introduction-FHF.tex
\section{Introduction}\label{sec:introduction}
\label{section:introduction}

Language Models (LMs), pre-trained on extensive datasets, have gained popularity in recent years. In Software Engineering (SE), domain-specific LMs, known as Code-LMs, trained on code, excel in various tasks such as code summarization, bug prediction, and method name prediction \cite{gu2022assemble,ahmed2022learning,zhang2022survey,guo2020graphcodebert,feng2020codebert,zhuo2024astraios,weyssow2023exploring,liu2023empirical}. Code-LMs are trained on code and/or its comments and fine-tuned on specific tasks.
Multiple studies explore various approaches to optimally fine-tune code-LMs, including considerations in using mono-lingual and multi-lingual datasets for fine-tuning~\cite{ahmed2021multilingual}. Others focus on enhancing results for low-resource languages~\cite{ahmed2022learning} and using alternative fine-tuning methods, such as Parameter Efficient Fine-Tuning (PEFT), to address computational resource constraints~\cite{goel2022cross, pfeiffer2020mad,pfeiffer2020adapterfusion,hu2021lora,rathnayake2022adapter,zhuo2024astraios,weyssow2023exploring,liu2023empirical}.
PEFT methods are introduced as alternatives to fine-tuning all parameters of code-LMs (i.e., full fine-tuning) as they require fewer parameters and computational costs~\cite{houlsby2019parameter}. 

PEFT approaches are attractive choices to fine-tune language models for downstream tasks with a smaller number of parameters both in NLP and SE \cite{pfeiffer2020adapterfusion,pfeiffer2020mad,saberi2023model,wang2023oneAdapter}. The importance of PEFT is more discernible when computational resources are limited. Additionally, it has been shown that language models do not perform well on low-resource languages (i.e., languages with {fewer training data}) \cite{feng2020codebert,guo2020graphcodebert}.

Previous research has shown that PEFT methods are effective in transferring natural language models to programming language tasks with reduced training costs~\cite{goel2022cross}. 
New PEFT architectures were also shown to be effective in imposing syntactical information to code-LMs and improving their results compared to fully fine-tuning them~\cite{saberi2023model}. 
Other works include empirical studies of PEFT in different settings, including in a multilingual setting, where a dataset of multiple programming languages is used to fine-tune the code-LM for downstream tasks~\cite{wang2023oneAdapter}, 
and evaluating the benefits of PEFT methods for SE tasks \cite{zhuo2024astraios,raffel2020exploring,weyssow2023exploring,liu2023empirical}. 

Although multiple studies have been conducted on the use of code-LMs in software engineering and on analyzing the benefits of PEFT in a multilingual setting, there is a gap in bridging the above-mentioned directions: transferring the knowledge from code-LMs of multiple programming languages to a target programming language task using PEFT.


Among the PEFT methods, a specific architecture known as \emph{AdapterFusion}~\cite{pfeiffer2020adapterfusion} learns to enhance the performance of a target task in a specific language, by leveraging similar latent information from other languages. 
AdapterFusion is based on \emph{Adapter}, a light-weight module inserted between the Transformer layers~ \cite{houlsby2019parameter}.
However, when we applied AdapterFusion in our experiments, we found that the model is learning mainly from the same language of the target task, rather than leveraging knowledge from other programming languages. 
As a result, we propose \textbf{Adversarial Fusion Adapter (AdvFusion)}, a new PEFT architecture that enforces AdapterFusion to first learn from the other programming languages before the target programming language. In this way, AdvFusion enhances the knowledge transfer among programming languages. 
The word `Adversarial' was chosen for AdvFusion {because we train the model in an adversarial manner, diverging from the conventional approach of creating adversarial samples.}

In evaluating AdvFusion, we first experiment with the effect of inserting mono-lingual PEFT methods trained on a specific task. We then use AdapterFusion to assess the existing PEFT architecture in learning from other programming languages. We compared our results against both mono-lingual and multi-lingual PEFT methods on Code-LMs.

We choose two target tasks in our experiments, code summarization and method name prediction, because they have been shown to enhance program comprehension and positively impact programmers' productivity~\cite{zhang2022survey}, and they are widely studied in the literature \cite{ahmed2021multilingual,hu2018summarizing,wang2023oneAdapter,ahmed2022learning,gu2022assemble}.


It is worth noting that we aim to use the knowledge transfer among programming languages through AdvFusion on smaller Code-LMs such as CodeT5+\cite{wang2023codet5+}, CodeBERT \cite{feng2020codebert} and GraphCoderBERT \cite{guo2020graphcodebert}, rather than Large Language Models trained on Code (Code-LLMs), such as CodeLlama~\cite{code-llama-roziere2023code} and StarCoder~\cite{li2023starcoder} with billions of parameters. 
Recent studies on how code-LMs and code-LLMs understand code, reveals that fine-tuning code-LMs on a target task could be more effective compared to code-LLMs~\cite{ma2024unveiling}. This is further supported by another work \cite{dou2023towards} for another SE task. Given these findings, we aim to exploit the capabilities of code-LMs and propose a faster fine-tuning approach to improve their performance.

Our evaluation on six programming languages from the CodeSearchNet dataset demonstrates that AdvFusion significantly enhances the performance of multilingual PEFT approaches in code summarization, showing improvements of up to 10\% across various models. Additionally, AdvFusion boosts method name prediction performance, achieving up to a 9\% increase in F1-score compared to AdapterFusion. Furthermore, AdvFusion outperforms LoRA, a widely recognized PEFT technique, by up to 12\% in code summarization and up to 32\% in method name prediction.
We observed that our proposed method also enhances the performance of low-resource languages in some cases during our experiments. 


Our contributions are as follows. 
\begin{itemize}
    \item Empirically study the capabilities of AdapterFusions for multilingual fine-tuning, in the context of code summarization and method name prediction.
    \item Propose an effective new PEFT architecture, AdvFusion, for fostering knowledge transfer from multiple programming languages to a target programming language task in the multi-lingual setting. 
    
\end{itemize}

The rest of the paper is organized as follows. In Section \ref{section:background}, we provide an overview of the necessary background information and then {discuss the effectiveness of the current PEFT method for multi-lingual training in section \ref{sec:motivatoin}.} We introduce our novel PEFT architecture, AdvFusion, in Section \ref{section:approach}. In Section \ref{section:experiment-setup} we provide the research questions and experimental setup. The results are explained in Section \ref{section:result} and they are discussed in more detail in Section \ref{section:discussion}. Sections \ref{section:related-work} and \ref{section:threat} are dedicated to the related works and threats to validity. Finally, we conclude the paper in Section \ref{section:conclusion}.

\begin{figure}
\centering
    \scalebox{.5}{
    \includegraphics{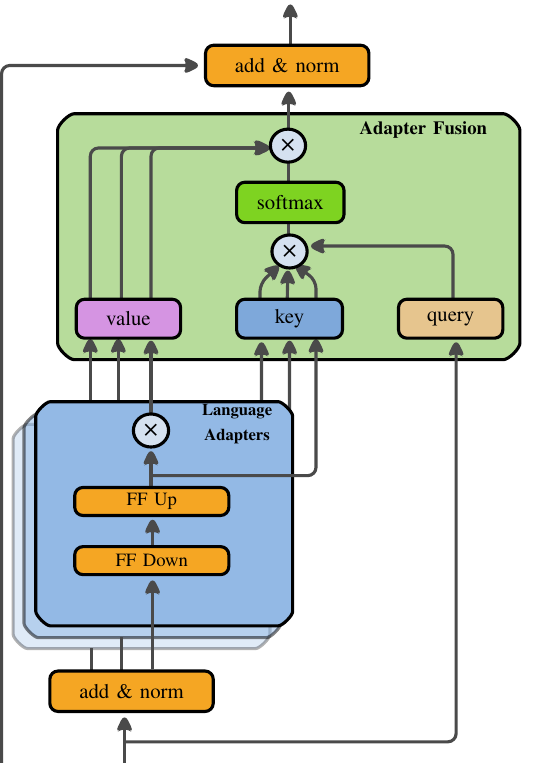}
    }
\caption{Internal structure of AdapterFusion. }

\label{fig:fusion-internal}
\end{figure}

%% file: Sections/background.tex
\section{Background}
\label{section:background}





\subsection{Adapters} 
\label{subsection:adapters}



We used various adapter types in our approach: task adapters, language adapters, and AdapterFusions. Adapters are lightweight modules added to a language model's internal structure, providing an efficient alternative to traditional fine-tuning for new tasks and preventing catastrophic forgetting. Adapters require less computational time and resources than fine-tuning.

Let $\Theta$ represent all weights of a pre-trained model. When an adapter $i$ is added, a new set of weights $\theta_i$ is created. During training, $\Theta$ remains frozen, and only $\theta_i$ is trained for the downstream task.

\subsubsection{Task Adapters}

The aim of a task adapter is to learn a task-specific functionality by training its weights on a target task dataset \cite{pfeiffer2020mad}. Task adapters consist of a  simple down- and up- projection combined with residual connections. Task adapter $TA_l$ at layer $l$ consists of a down-projection $D \in R^{h\times d} $  where $h$ is the hidden size of the Transformer and $d$ is the dimension of the adapter. The down-sampled representations are then fed to a ReLU activation followed by an up-projection transformation $U \in R^{d\times h} $ at each layer. This is shown in Equation \ref{eq:task-adapter}:

\begin{equation}
TaskAdapter_l(h_l,r_l)=U_l(ReLU(D_l(h_l)))+r_l
\label{eq:task-adapter}
\end{equation}



\subsubsection{Language Adapters}
Language adapters learn language-specific features by training their weights on an abstract objective function such as MLM \cite{pfeiffer2020mad}. The language adapter $LA_l$ at layer $l$ has the same architecture as a task adapter. The internal structure of a language adapter consists of a down-projection $D \in R^{h\times d} $ with a ReLU activation, followed by an up-projection $U \in R^{d\times h} $, as shown in Equation \ref{eq:language-adapter}:

\begin{equation}
LanguageAdapter_l(h_l,r_l)=U_l(ReLU(D_l(h_l)))+r_l
\label{eq:language-adapter}
\end{equation}

where $h_l$ and $r_l$ are defined similar to task adapters. 
Language adapters differ from task adapters in that they are trained on unlabeled data using Masked Language Modeling (MLM), focusing on learning specific language embeddings. These embeddings can then be employed as input for task adapters or combined with AdapterFusion for extracting latent knowledge for downstream tasks. 

\subsubsection{AdapterFusion}
Language adapters are introduced to extract language-specific embeddings from the internal structure of an LM based on an abstract objective function such as MLM to learn the general representations of a language. AdapterFusion aims to extract and compose the latent knowledge from multiple language adapters for a downstream task such as code summarization. For example, given a set of $N$ language adapters, the output of adapterFusion is a weighted sum of outputs from the language adapters, while the weights of the LMs weights ( $\Theta$) and the language adapter $(\theta_1,...,\theta_N)$ are fixed. This is shown in Equation \ref{eq:fusion-adapter}:
\begin{equation}
\Phi= \textrm{argmin } L(D;\Theta,\theta_1,...,\theta_N)
\label{eq:fusion-adapter}
\end{equation}

where $\Phi$ consists of the $Key_l$, $Value_l$ and $Query_l$ metrics at each layer $l$. At each Transformer block, the output of the feed-forward sub-layer is taken to be the $Query$, and the output of each language adapter is used for both $Key$ and $Value$ vectors.
Fig.~\ref{fig:fusion-internal} shows the internal structure of AdapterFusion.








%% file: Sections/Motivation.tex
\section{Effectiveness of AdapterFusion} \label{sec:motivatoin}

Previous research \cite{ahmed2021multilingual} shows that fully fine-tuning a Code-LM on a multilingual dataset improves the performance of the models for code summarization task. 
Additionally, it is shown that during the fine-tuning phase, if we encourage the weights to stay closer to the initial weights of the LMs, we can avoid catastrophic forgetting and stabilize the fine-tuning process~\cite{lee2019mixout}. 
Therefore, we hypothesize that using a PEFT architecture that is designed to learn from other tasks/languages, i.e., AdapterFusion, will improve the performance compared to fully fine-tuning a Code-LM on multilingual data. 
To analyze the effectiveness of AdapterFusion in a multilingual setting, we choose three Code-LMs, CodeT5+(220M)\cite{wang2023codet5+} denoted as CodeT5p for the rest of the paper, CodeBERT~\cite{feng2020codebert} and GraphCodeBERT~\cite{guo2020graphcodebert}.

We first train language adapters, one for each of the six languages from the CodeSearchNet dataset. Then, we stack the trained language adapters and fine-tune AdapterFusion for two tasks, code summarization and method name prediction, fixing the parameters of the Code-LMs and the language adapters. We denote these models as CodeT5p+AdapterFusion, CodeBERT+AdapterFusion and GraphCodeBERT+AdapterFusion. 

Additionally, we performed experiments in a monolingual setting to evaluate the effectiveness of multilingual versus monolingual PEFT approaches. In this setting, the fine-tuning dataset consists of code summarization or method name prediction tasks for a single programming language. For these experiments, we fine-tuned the selected code language models (Code-LMs) using two types of adapters: task adapters and LoRA adapters. To do this, the base weights of the code models were kept frozen, while we integrated the corresponding adapters into the models for task-specific fine-tuning. We refer to these modified models as CodeT5p+TaskAdapter, CodeBERT+TaskAdapter, and GraphCodeBERT+TaskAdapter when using task adapters, and CodeT5p+LoRA, CodeBERT+LoRA, and GraphCodeBERT+LoRA when using LoRA adapters.

\begin{figure}

\includegraphics[width=\columnwidth]{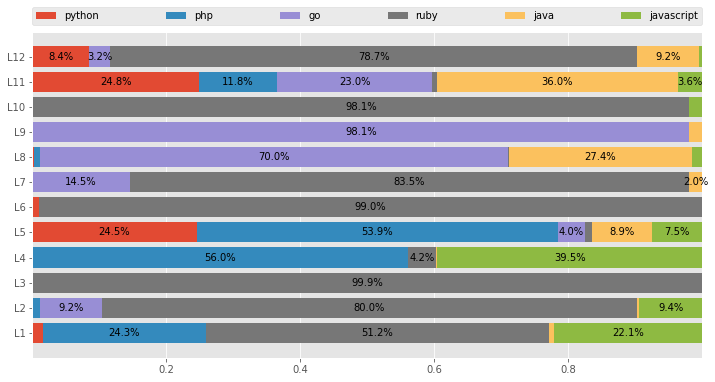}
\centering
\caption{The attention contribution from each programming language at each layer when we feed the Ruby test dataset to the fine-tuned AdapterFusion model.}
\label{fig:fusion-ruby}
\end{figure}

The obtained scores for code summarization are shown in Table~\ref{table:different_methods}. There are several observations here. First, the use of TaskAdapters and LoRA in conjunction with Code-LMs leads to improved performance compared to full fine-tuning for both CodeBERT and GraphCodeBERT on three specific languages: Ruby, JavaScript, and Go. Additionally, the results achieved using TaskAdapters and LoRA are comparable to those obtained by other models in these languages.

In particular, we employed only $0.89$ million trainable parameters for fine-tuning with task adapters, significantly fewer than the fully fine-tuned CodeBERT and GraphCodeBERT models with $110$ million parameters. This result is aligned with previous research \cite{saberi2023utilization}. 

Second, we noticed that AdapterFusion helps improve the results of mono-lingual adapters, i.e., Code-LM+TaskAdapters for CodeBERT, and has similar scores for GraphCodeBERT and CodeT5p. 

Table~\ref{table:mnp} presents the results for method name prediction.
For CodeBERT and GraphCodeBERT, AdapterFusion demonstrates superior performance compared to TaskAdapters and LoRA. However, in the case of CodeT5p, the impact of the multilingual setting is less pronounced. We hypothesize that this is because CodeT5p, being a larger model, can effectively leverage the CodeSearchNet dataset during pre-training, reducing the need for further adaptation in multilingual settings.


Here, we aim to uncover the underlying mechanics of applying AdapterFusion in a multilingual PEFT setting, providing insight into its inner workings and effects on model performance. Figure~\ref{fig:fusion-ruby} presents the attention contribution of each programming language when AdapterFusion is used for code summarization in Ruby. 
The x-axis shows the attention score and the y-axis shows the percentage contribution of each language at each layer. The color bars show the contributions from each of the six programming languages. In most layers, a high percentage of attention (more than 80\%) is towards Ruby (the gray bar). This shows that AdapterFusion tends to pay more attention to the language adapter that corresponds to the same language of the target task.
This figure also suggested that AdapterFusion could benefit from an architectural change to achieve its goal, which is to learn from other programming languages. 
In the next section, we detail our proposed architecture, AdvFusion to achieve this goal.

%% file: Sections/AdversarialFusion.tex
\section{Adversarial Fusion Adapter}
\label{section:approach}
In this section, we describe the architecture of our approach, AdvFusion, before proposing a learning algorithm for it. 

\subsection{Architecture}
\label{subsec:architecture}

\begin{figure}
\centering
    \scalebox{.5}{
    \includegraphics{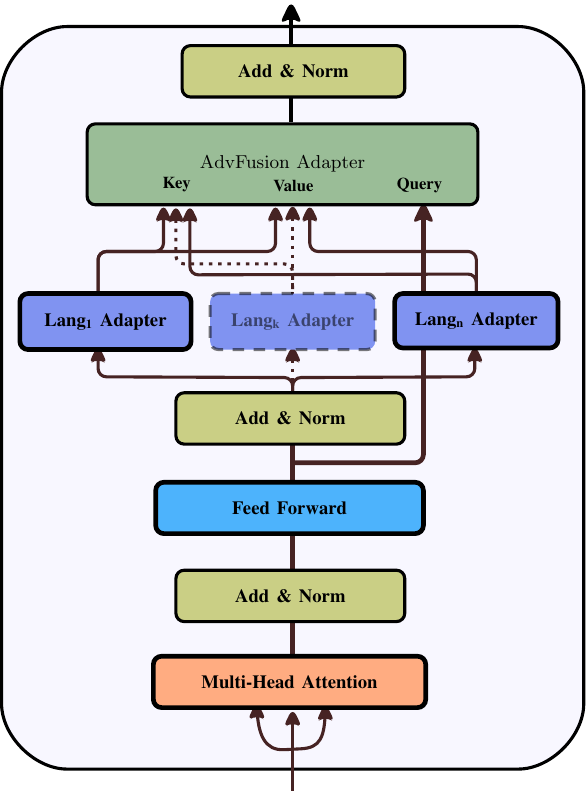}
    }
\caption{The adversarial training phase of AdvFusion. 
}
\label{fig:advfusionphase1}

\end{figure}

\begin{figure}
\centering
    \scalebox{.5}{
    \includegraphics{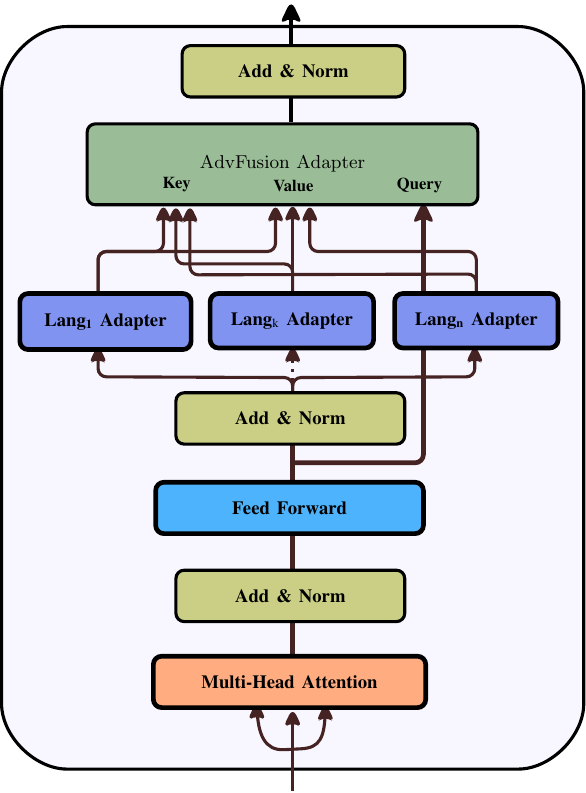}
    }
\caption{ The fine-tuning phase of AdvFusion. }


\label{fig:advfusionphase2}

\end{figure}

AdapterFusion can leverage the language adapter corresponding to the language of the current input better \cite{pfeiffer2020adapterfusion}, i.e., it pays more attention to the language adapter of the target task. This is mainly due to its internal attention mechanism.  
This mechanism prevents the effective utilization of the other language adapters, thus rendering them redundant. 
In light of this, we propose a new architecture, AdvFusion, that requires AdapterFusion to learn more from the other language adapters that are trained using a different language from the target task. 
Our approach consists of two training phases, the Adversarial training phase and the Fine-tuning phase: 
\begin{enumerate}

    \item 
    Adversarial training phase (see Fig.~ \ref{fig:advfusionphase1}): In this phase, (i) the weights of the language adapter that corresponds to the language of the target task are set to zero, while (ii) the weights of the code-LM and the other language adapters are fixed. Then, (iii) AdvFusion is trained on the entire dataset. This phase allows AdvFusion to learn from the other language adapters.
    \label{phase1}
    \item
    Fine-tuning phase (see Fig.~\ref{fig:advfusionphase2}): In this phase, AdvFusion would have learnt from the other language adapters in the earlier phase. However, we still want AdvFusion to learn from the language adapter that corresponds to the language of the target task. Thus, (i) we restore the trained weights of the language adapter that corresponds to the language of the target task, while still (ii) fixing the weights of the code-LM and all language adapters. Then, (iii) the weights of AdvFusion are fine-tuned.
    \label{phase2}

\end{enumerate}


\subsection{Learning Algorithm}
\label{subsec:learning}
In this section, we formalize the learning procedure of AdvFusions. Let $\Theta$ and $\theta_i$ denote the parameters of the code-LM and each language adapter, $language_i$, respectively. We introduce the $ \Psi$ parameters to learn an embedding space from $N$ language adapters for a downstream task. For the adversarial training phase, we formalize it as follows:
\begin{equation}
\begin{aligned}
\Psi \leftarrow \mathop{\textrm{argmin }}_{\Psi}  
 \sum_{m=1}^{N} L(D_{m}; \Theta, \theta_{1},..,\theta_{m-1},\theta_{m+1},..,\theta_{N},\Psi)
\end{aligned}
\label{eq:adv-phase1}
\end{equation}

where L is the loss function of the downstream task, and $D_m$ denotes the $language_m$ dataset. In this step, AdvFusion learns to compose the embeddings of $N-1$ language adapters at each training step (recall that we are only interested in learning from the other language adapters that differ from the language of the target task in the adversarial training phase, thus we are only learning from $N-1$ language adapters). 

In the second phase, we employ all the language adapters to train the $\Psi$ parameters as follows:

\begin{equation}
\begin{aligned}
\Psi \leftarrow \mathop{\textrm{argmin }}_{\Psi}  
 \sum_{m=1}^{N} L(D_{m}; \Theta, \theta_{1},..,\theta_{N},\Psi)
\end{aligned}
\label{eq:adv-phase2}
\end{equation}

As illustrated in Fig. \ref{fig:advfusionphase1}, $ \Psi$ consists of the \textit{Key}, \textit{Value} and \textit{Query} parameters, denoted by $K_l$, $V_l$ and $Q_l$ at the Transformer layer $l$, respectively.

Let $h_l$ denote the output of the feed-forward sub-component at the Transformer layer $l$. This is an input to AdvFusion. The output of the language adapter $i$ at the Transformer layer $l$, denoted as $z_{l,i}$, is the input for both the \textit{Key} and \textit{Value} transformations at the Transformer layer $l$. We compute the output of AdvFusion, denoted by $O_l$, as follows:

\begin{equation}
\begin{aligned}
& S_{l} = \textrm{softmax}(h^T_l Q_l \otimes z^T_{l,n} K_l) \\
& z'_{l,n} = z^T_{l,n} V_l \\
& Z'_l = [z'_{l,0},...,z'_{l,N}] \\
& O_l = S^T_l Z'_l
\label{eq:adv-phase-output}
\end{aligned}
\end{equation}


Given the embeddings of each language adapter ($z_n$), AdvFusion learns a weighted mixer of the available trained language adapters. In equation \ref{eq:adv-phase-output}, $\otimes$ represents the dot product and $n$ refers to two different things in each of the phases in AdvFusion. In the adversarial training phase (first phase),  $n \in \{1,...,m-1,m+1,...,N\}$ while in the fine-tuning phase (second phase), $n \in \{1,...,N\}$.

%% file: Sections/ResearchQuestions.tex
\section{Experiment Setup} \label{section:experiment-setup}
In this section, we discuss the research questions and the details of our study design and experiment setup. 

\subsection{Research Questions}

In this study, we conduct experiments to answer the following research questions:

\textbf{RQ1:} 
\textit{{Does using AdvFusion lead to a performance improvement in multilingual fine-tuning? }}


In this RQ, our goal is to evaluate the effectiveness of multilingual fine-tuning a language model designed for code, utilizing our innovative approach, AdvFusion. The primary objective of AdvFusion is to emphasize on knowledge transfer across various programming languages, ultimately improving the overall performance and adaptability of the model.
AdvFusion operates by selectively extracting, exploiting, and aggregating the latent knowledge acquired from different programming languages. Within the scope of this RQ, we aim to assess the extent to which the programming languages can benefit from the knowledge derived from others. 
As this learning is enforced in AdvFusion, it is compared against  AdapterFusion. 

\textbf{RQ2: }
\textit{How much attention is placed on the target language from the other (non-target) languages in AdvFusion?}

Constructing good-quality datasets for target programming languages is a time-consuming and challenging task. Previous studies have attempted to take advantage of other (non-target) programming languages for a target programming language through a multilingual training dataset \cite{ahmed2021multilingual,chen2022transferability}.
As the AdvFusion architecture enables learning from other languages in an initial stage, we are interested to see the amount of contributed attention from other languages. 
In this RQ, we use attention analysis to evaluate how much we can learn from other programming languages for a target language.
More specifically, we calculate the participation of each language at every transformer block of the model by measuring the percentage of attention we get from each language adapter for each sample over a target language dataset.

\subsection{Backbone Models} \label{sec:backbones}

We have selected CodeT5+(220M)\cite{wang2023codet5+}, CodeBERT\cite{feng2020codebert} and GraphCodeBERT\cite{guo2020graphcodebert}  as our baseline models for several reasons. CodeT5+220M was introduced recently and is considered an improved version of CodeT5\cite{wang2021codet5}. The other models have been extensively researched in the field of software engineering~\cite{ahmed2021multilingual,saberi2023model,lu2021codexglue,wang2023oneAdapter,chung2014empirical}.
Additionally, these models are studied for multilingual fine-tuning for these two tasks and therefore serve as a basis in our comparisons \cite{ahmed2021multilingual}. 



{Please note that recent studies by \cite{ma2024unveiling} on how code language models (code-LMs) and code large language models understand code reveal that fine-tuning smaller models on a target task could be more effective as compared to code-LLMs with billions of parameters. This finding is further supported by \cite{dou2023towards} for another software engineering task. Given these findings, we aim to exploit the capabilities of code-LMs and propose a speedier fine-tuning approach to improve their performance.}



\textbf{CodeT5+} is an advanced code-LM designed by \cite{wang2023codet5+} to overcome limitations in existing code models, which often rely on rigid encoder-only or decoder-only architectures. It introduces a flexible, modular approach, allowing customization for various code-related tasks. CodeT5+ achieves superior performance compared to other code-LMs of similar size by incorporating a combination of pre-training techniques, including span denoising and contrastive learning.

\textbf{CodeBERT}, as introduced by \cite{feng2020codebert}, is a bimodal pretrained model designed for both natural language and programming language understanding. Its architecture is based on Transformers. CodeBERT employs two pre-training objectives, namely Masked Language Modeling and Replaced Token Detection. These objectives are specifically chosen to enhance its capabilities in supporting tasks such as code search and code documentation generation.

\textbf{GraphCodeBERT}, introduced by \cite{guo2020graphcodebert}, is a pioneering pre-trained model designed to enhance code comprehension tasks such as code summarization. GraphCodeBERT utilizes semantic-level information from code, specifically focusing on aspects like data flow. This pre-training approach employs a 12-layer transformer-based architecture. It is pre-trained on Masks Language Modeling, Edge Prediction and Node Alignment objective functions.

\subsection{Tasks}
\subsubsection{Code Summarization} \label{sec:task1}

Given a code snippet, the task of code summarization is to describe its functionality. 
{It enhances code readability, aids in program comprehension, and facilitates easier maintenance and documentation. By providing summaries, developers can quickly understand the purpose and functionality of a piece of code without delving into its implementation details \cite{nie2022impact}.}
Code summarization is chosen as it is a widely studied task and the effects of multilingual fine-tuning for this task is investigated in previous research \cite{ahmed2021multilingual,wang2023oneAdapter}. 



\textbf{Evaluation metric}
We evaluate the code summarization task using smooth-BLEU-4 \cite{papineni2002bleu}, which is a widely used metric in natural language generation tasks and many software engineering studies \cite{feng2020codebert,guo2020graphcodebert,wang2021codet5,wang2023oneAdapter,tang2020multilingual}.
BLEU is a precision-based metric that measures the n-gram geometric precision between the generated summary (i.e., n-gram hit) and the ground truth summary (i.e., total n-gram count) \cite{papineni2002bleu}.

\subsubsection{Method Name Prediction} \label{sec:task2}

 The objective of the method name prediction task is to generate the most fitting method name that describes the purpose and functionality of the method's code. {This task is chosen because naming methods accurately is crucial for code readability, maintainability, and understanding.} 
 
 


\textbf{Evaluation metric}
We report precision, recall and F1-score for the generated method names. F1 Score is the weighted average of Precision and Recall: $F1 = \frac{2 \cdot (P \cdot R)}{P + R}$. 
Where P and R stand for Precision and Recall, respectively.

Precision is computed as $P = \frac{TP}{TP+FP}$, whereas Recall is calculated as $R = \frac{TP}{TP+FN}$. P is calculated as the length of the intersection of ground truth tokens and generated output tokens (i.e., TP) divided by the length of output tokens (i.e., TP + FP). Similarly, R represents the recall, calculated as the length of the intersection of ground truth tokens and generated output tokens (i.e., TP) divided by the number of ground truth tokens (i.e., TP + FN).


\subsection{Baselines}

AdvFision in a Code-LM should be compared against the same Code-LM+AdapterFusion. For example, we should compare CodeBERT+AdvFusion with CodeBERT+AdapterFusion. 

To show the effectiveness of the base PEFT architecture we used, we also provide the results for mono-lingual fine-tuning, including Code-LMs+TaskAdapters and Code-LMs+LoRA~\cite{hu2021lora}. 
{Note that we perform experiments on LoRA \cite{hu2021lora} as it is a widely used PEFT method. This enables us to compare its performance against other approaches and AdvFusion.}




\subsection{Experiments Design}


Our study delves into code summarization and method name prediction tasks, aiming to gather diverse programming language knowledge. 



To train AdvFusion, in the first phase, we (1) fix the weights of the language adapter, (2) temporarily set the weights of the language adapter corresponding to the current input (i.e., the language of the target task) to zero, and (3) train the weights of AdvFusion on our target task. In the second phase, we restore the weights of the language adapter corresponding to the input and allow AdvFusion to learn from the language adapter that corresponds to the language of the current input.


For \textit{RQ2}, we evaluate the contribution of each programming language for a target programming language. Here, we choose Ruby, as our experiments in Section~\ref{sec:motivatoin} show that it could benefit from other programming languages. Ruby is also used as a low-resource language in previous studies~\cite{chen2022transferability} and it has been shown that it can benefit from other languages. 


We compute the contributions by feeding the Ruby test dataset into CodeBERT+AdapterFusion. Then, we aggregate the attention scores from each language adapter in each layer, normalize them (i.e., min-max normalization), and obtain the percentage of each language's contribution. We repeat these steps for CodeBERT+AdvFusion to compare its ability with AdapterFusion in extracting knowledge from other programming languages for Ruby. You can find the other language contributions on the repository page\footnote{https://anonymous.4open.science/r/AdvFusion-5841/README.md}. All experiments are conducted on an Nvidia Tesla V100 32GB GPU.

\subsection{Training Details}



As Pfeiffer et al. have performed an extensive hyperparameter search over adapters, we use their reported optimal settings in our adapters' hyperparameters \cite{pfeiffer2020adapterfusion}.

\label{subsec:dataset}
 We use the CodeSearchNet datasets \cite{husain2019codesearchnet} for training the language adapters. It consists of datasets from 6 programming languages and the size of each language is shown in Table \ref{table:plm}. We train language adapters using Mask Language Modelling.

{We fine-tune AdapterFusion and AdvFusion adapters on the CodeSearchNet datasets using the next token prediction objective function for
code summarization.}
{For the method name prediction task, we exclusively utilize the code portion of the CodeSearchNet datasets. We then mask the method names
and let each approach suggest new method
names using the next token generation objective function.}

\begin{table}[h!]
\centering
   
    \begin{tabular}{|c | c | c| c|} 
        \hline
        Language & \# of Bimodal Data & Language & \# of Bimodal Data \\ [0.5ex] 
        \hline
         Ruby  & 24,927  & Python  & 251,820\\
         JavaScript  & 58,025  & Java  & 164,923\\
         Go  & 167,288  & PHP  & 241,241 \\ 
         \hline
    \end{tabular}
    \caption{Dataset statistics.  \cite{husain2019codesearchnet}}
    \label{table:plm}
  
\end{table}

%% file: Sections/Results.tex
\section{Results} \label{section:result}

In this section, we present the results of our experiments, including the time reduction when AdvFusion is used.

\begin{table}
    \centering
    \scalebox{.95}{
    \begin{tabular}{c|c|c|c}
    \hline
         Language & CodeBERT & CodeBERT+AdvFusion & Time reduction\\
         \hline
         Ruby &  492 & 328 & -33\% \textcolor{Gr}{$\downarrow$}\\ 
         JavaScript &  493 & 344 & -30\% \textcolor{Gr}{$\downarrow$}\\ 
         Go  &   511 & 336 & -34\% \textcolor{Gr}{$\downarrow$}\\ 
         Python  & 493 & 323 & -34\% \textcolor{Gr}{$\downarrow$}\\ 
         Java &  494 & 341 & -31\% \textcolor{Gr}{$\downarrow$}\\ 
         PHP &  506 & 338 & -33\% \textcolor{Gr}{$\downarrow$}\\
         \hline
    \end{tabular}
    }
    \caption{AdvFusion time efficiency for code summarization. Numbers represent training time in minutes, with the last column showing percentage improvement. Times reflect training for 20,000 training steps.}
    \label{table:adv-time}
\end{table}


\begin{table*}[t!]
\centering

\begin{tabular}{|l | c | c| c| c| c| c|} 
 \hline
 \textbf{Models} & \textbf{Ruby} & \textbf{JavaScript} & \textbf{Go} & \textbf{Python} & \textbf{Java} & \textbf{PHP} \\ [0.5ex] 
  \hline

 CodeT5p+AdvFusion & 14.70 & \textbf{14.96}  & 18.25  & \textbf{18.98} & \textbf{18.78}  & \textbf{23.87}  \\ 
 CodeT5p+AdapterFusion & \textbf{14.79} & 14.82  & 18.30 & 18.94  & 18.71  & 23.80 \\ 

 CodeT5p+TaskAdapter & 13.99 & 14.31  & \textbf{18.34} & 18.91  & 18.68  & 23.71 \\ 

 CodeT5p+LoRA & 13.56 & 14.25  & 18.08 & 18.88 & 18.67  & 23.47 \\ 
\rowcolor{gray!30}
  CodeT5p (Full Fine-Tuned)  & 14.55 & 15.16  & 19.00 & 19.77 & 19.60  & 25.13 \\ 
 \hline
 
 GraphCodeBERT+AdvFusion & \textbf{16.47} & \textbf{15.89}  & \textbf{19.96}  & 18.49 & \textbf{18.97}  & \textbf{24.83}  \\ 
 GraphCodeBERT+AdapterFusion & 15.57 & 14.49  & 18.21 & 17.86  & 18.21  & 23.54 \\ 

 GraphCodeBERT+TaskAdapter & 14.39 & 14.53  & 18.47 & 17.88  & 17.29  & 23.36 \\ 

 GraphCodeBERT+LoRA & 14.48 & 14.63  & 17.8 & \textbf{18.50}  & 17.16  & 24.13 \\ 

\rowcolor{gray!30}
GraphCodeBERT (Full Fine-Tuned)  & 12.62 & 14.79 &18.40 & 18.02 & 19.22 & 25.45\\ 
 \hline
 CodeBERT+AdvFusion & \textbf{16.53} & \textbf{16.80} & \textbf{19.69} & 18.28 & \textbf{19.94} &25.20  \\ 
 CodeBERT+AdapterFusion & 15.38 & 15.88 & 18.31 & 18.40 & 19.04 & 25.17 \\ 

 CodeBERT+TaskAdapter & 14.12 & 15.67 &18.51 & \textbf{18.47} & 18.99 & \textbf{25.55} \\ 

 CodeBERT+LoRA & 12.27 & 13.67  &19.01 & 17.07 & 16.58 & 23.08 \\ 
 \rowcolor{gray!30}
 CodeBERT(Full Fine-Tuned) & 12.16 & 14.90 &18.07 & 19.06 & 17.65 & 25.16 \\ 
\hline



 \hline
\end{tabular}

\caption{Smooth BLEU-4 scores on code summarization. When AdvFusion is combined with Code-LMs, we saw an improved performance in the majority of the datasets.
}
\label{table:different_methods}
\end{table*}

\begin{table*}[!ht]
\centering
\begin{adjustbox}{width=1\textwidth,center}
\begin{tabular}{@{}cccccccccccccccccccccccc@{}}
 \hline
\textbf{Model} &\multicolumn{3}{c}{\textbf{Ruby}}  &  & \multicolumn{3}{c}{\textbf{Javascript}} &  & \multicolumn{3}{c}{\textbf{Go}} &  & \multicolumn{3}{c}{\textbf{Python}} &  & \multicolumn{3}{c}{\textbf{Java}} &  & \multicolumn{3}{c}{\textbf{PHP}}\\ \cline{2-4}\cline{6-8} \cline{10-12}\cline{14-16} \cline{18-20} \cline{22-24}
&  $P$  & $R$ & $F1$ &  & $P$ & $R$ & $F1$ &  & $P$ & $R$ & $F1$ &  & $P$ & $R$ & $F1$ &  & $P$ & $R$ & $F1$ &  & $P$ & $R$ & $F1$  \\ [1em] 

\specialcell{CodeT5p + \\AdvFusion} &  \textbf{0.55}   &   \textbf{0.55}    &   \textbf{0.55}    &  &   \textbf{0.59}    &   \textbf{0.56}  &   \textbf{0.58}    
& &  \textbf{0.58}   &   \textbf{0.56}  &  \textbf{0.57}   &  &   0.61  &  0.61  &  0.61  & &  \textbf{0.61}   & 0.57   &  \textbf{0.59}   & &  \textbf{0.49}  &  0.46  &  \textbf{0.48} \\ [1em]

\specialcell{CodeT5p + \\AdapterFusion} &  0.54   &   0.54    &   0.54    &  &   0.57    &   0.55  &   0.56    
& &  0.55   &   0.53  &  0.54   &  &   0.60  &  0.59  &  0.60  & &  0.59   & 0.56   &  0.57   & &  0.47  &  0.44  &  0.46 \\ [1em]

\specialcell{CodeT5p + \\TaskAdapter} &  0.53   &   0.54    &   0.54    &  &   0.54    &   0.57  &   0.55    
& &  0.55   &   0.57  &  0.56   &  &   0.61  &  0.61  &  0.61  & &  0.60   & 0.57   &  0.59   & &  0.48  &  0.46  &  0.47 \\ [1em]

CodeT5p+LoRA &  0.53   &   0.52    &   0.53    &  &   0.53    &   0.56  &   0.55    
& &  0.54   &   0.56  &  0.55   &  &   \textbf{0.61}  &  \textbf{0.61}  &  \textbf{0.61}  & &  0.57   & \textbf{0.59}   &  0.58   & &  0.48  &  0.45  &  0.46 \\ [1em]

\hline

\specialcell{CodeBERT + \\AdvFusion} &  \textbf{0.39}   &   0.32    &   \textbf{0.35}    &  &   0.19    &   0.14  &   0.16    
& &  \textbf{0.46}   &   \textbf{0.46}  &  \textbf{0.45}   &  &   \textbf{0.47}  &  \textbf{0.45}  &  \textbf{0.46}  & &  0.43   & 0.34   &  0.37   & &  \textbf{0.45}  &  \textbf{0.43}  &  \textbf{0.44} \\ [1em]

\specialcell{CodeBERT + \\AdapterFusion} &  0.38   &   0.30    &   0.32    &  &   0.19    &   0.14  &   0.16    
& &  0.45   &   0.40  &  0.41   &  &   0.44  &  0.34  &  0.37  & &  0.43   & 0.34   &  0.37   & &  0.45  &  0.38  &  0.40 \\ [1em]

\specialcell{CodeBERT + \\TaskAdapter} &  0.35   &   0.30    &   0.30    &  &   0.19    &   0.14  &   0.16    
& &  0.45   &   0.40  &  0.41   &  &   0.44  &  0.34  &  0.37  & &  0.43   & 0.34   &  0.37   & &  0.45  &  0.38  &  0.40 \\ [1em]

CodeBERT+LoRA &  0.36   &   \textbf{0.33}    &   0.34    &  &   \textbf{0.21}    &   \textbf{0.16}  &   \textbf{0.18}    
& &  0.45   &   0.42  &  0.43   &  &   0.43  &  0.45  &  0.44  & &  0.42   & \textbf{0.40}   &  \textbf{0.41}   & &  0.41  &  0.44  &  0.43 \\ [1em]

\hline

 \textbf{\specialcell{ Graph\\CodeBERT + \\AdvFusion}} &  \textbf{0.42}   &   \textbf{0.32}    &   \textbf{0.36}    &  &   \textbf{0.58}    &   \textbf{0.58}  &   \textbf{0.58}   
& &  \textbf{0.51}   &   \textbf{0.51}  &  \textbf{0.51}   &  &   0.49  &  0.40  & 0.42  & &  \textbf{0.52}   &  \textbf{0.50}   &  \textbf{0.51}   & &  \textbf{0.54}  &  \textbf{0.53}  &  \textbf{0.54} \\[1em]

 \textbf{\specialcell{ Graph\\CodeBERT + \\AdapterFusion}} &  0.40   &   0.30    &   0.35    &  &   0.57    &   0.57  &   0.57   
& &  0.48   &   0.49  &  0.47   &  &   0.48  &  0.38  & 0.41  & &  0.48   &  0.49   &  0.48   & &  0.52  &  0.50  &  0.51 \\[1em]

 \specialcell{Graph\\CodeBERT + \\TaskAdapter} &  0.40   &   0.33    &   0.35    &  &   0.24    &   0.22  &   0.23    
& &  0.47   &   0.42  &  0.43   &  &   0.47  &  0.38  & 0.40  & &  0.45   &  0.37   &  0.40   & &  0.48  &  0.41  &  0.43 \\[1em]

 \specialcell{Graph\\CodeBERT+LoRA} &  0.39   &   0.32    &   0.35    &  &   0.28    &   0.24  &   0.26    
& &  0.51   &   0.45  &  0.47   &  &   \textbf{0.50}  &  \textbf{0.44}  & \textbf{0.45}  & &  0.48   & 0.43   &  0.44  & &  0.49  &  0.45  &  0.46 \\[1em]

 \bottomrule
\end{tabular}
\end{adjustbox}
\caption{
The Precision (P), Recall (R), and F1-Score (F1) metrics were assessed on each programming language across various settings. When AdvFusion is combined with Code-LMs, we saw an improved performance in majority of the datasets.}
\label{table:mnp}
\end{table*}

\begin{table*}[h!]
\centering
\begin{adjustbox}{width=1\textwidth,center}
\begin{tabular}{|c|c|c|c|}
\hline
\textbf{Samples} & \textbf{CodeT5p+Fusion}  & \textbf{CodeT5p+AdvFusion} & \textbf{Target} \\ \hline
sample 1(Javascript) & Parse a segment \colorbox{red!30}{of a string}. & Parse a segment into \colorbox{green!30}{a single object}. & Parse a segment and convert it into json. \\ \hline

sample 2(Javascript) & Transform a metadata object into a \colorbox{red!30}{ string}. & \begin{tabular}{@{}c@{}} Transform a metadata object \\  into a \colorbox{green!30}{list of tokens}. \end{tabular}  & \begin{tabular}{@{}c@{}} Transform token names to formats expected \\ by Sassdoc for descriptions and aliases \end{tabular} \\ \hline

sample 3(PHP) & Create a new \colorbox{red!30}{model}. & Create a new \colorbox{green!30}{database table}. & Store new database table. \\ \hline

sample 4(PHP) & \colorbox{red!30}{Deletes all files} in the media picker table. & \colorbox{green!30}{Cleanup}  the media picker data & \begin{tabular}{@{}c@{}} Remove translations \\ images and files related to a BREAD item.\end{tabular} \\ \hline

sample 5(Go) & \begin{tabular}{@{}c@{}} Percentiles returns the percentage of \\ the given \colorbox{red!30}{number of elements}. \end{tabular}&  \begin{tabular}{@{}c@{}} Percentiles returns the percentage \\ of the given \colorbox{green!30}{array of floats}. \end{tabular}& \begin{tabular}{@{}c@{}}Percentiles returns percentile \\ distribution of float64 slice.\end{tabular} \\ \hline

sample 6(Go) & newPipelineHandler creates a new \colorbox{red!30}{pipeline handler}. & \begin{tabular}{@{}c@{}} newPipelineHandler returns a new \colorbox{green!30}{http}\\ \colorbox{green!30}{Handler that will handle the pipeline request}. \end{tabular} & \begin{tabular}{@{}c@{}c@{}c@{}} newPipelineHandler returns a handler for handling \\ raft messages from pipeline for RaftPrefix.\\ The handler reads out the raft message from request body \\and forwards it to the given raft state machine for processing.\end{tabular} \\ \hline
\end{tabular}
\end{adjustbox}

\caption{Comparison between CodeT5p+Fusion and CodeT5p+AdvFusion outputs with their ground truth. Samples are selected from the test set results of CodeSearchNet dataset.}
\label{table:manual_comparison}
\end{table*}

\subsection{RQ1: Performance of Multilingual PEFT with AdvFusion}

In this RQ, we evaluate how much improvement we could gain by using other programming languages, therefore, transferring knowledge in the multilingual parameter efficient fine-tuning of Code-LMs.


In Table \ref{table:different_methods}, we present the BLEU scores for both multilingual and monolingual PEFT approaches applied to Code-LMs. The multilingual approaches include Code-LM with AdvFusion and AdapterFusion, while the monolingual approaches involve Code-LM with TaskAdapter and LoRA. Although the base Code-LMs are the same, the key difference lies in the fine-tuning strategies used.

With CodeBERT+AdvFusion, we observe improvements of 8\%, 6\%, 7\%, and 5\% in BLEU scores for Ruby, JavaScript, Go, and Java, respectively. Similarly, with GraphCodeBERT+AdvFusion, we see gains of 6\%, 10\%, 8\%, and 4\% for the same languages. However, for Python and PHP, CodeBERT+TaskAdapter and GraphCodeBERT+LoRA show higher performance. We attribute this to the larger training data available for Python and PHP compared to Ruby and JavaScript, which have fewer samples. The smaller datasets for Ruby and JavaScript suggest that these languages still benefit from additional knowledge transfer.

We also compare the performance of AdvFusion with the state-of-the-art PEFT method, LoRA. In five of the programming languages evaluated (excluding Python), AdvFusion consistently outperforms LoRA. Performance gains are especially pronounced for CodeBERT and GraphCodeBERT, while the improvement for CodeT5p is less substantial. To better understand this discrepancy, we manually analyzed the outputs of CodeT5p+AdapterFusion and CodeT5p+AdvFusion against the ground truth targets, as shown in Table \ref{table:manual_comparison}. Although the overall improvement for CodeT5p is modest, our analysis reveals that AdvFusion tends to capture finer details more effectively.

In terms of parameter efficiency, both AdapterFusion and AdvFusion are more efficient than fully fine-tuning CodeBERT.
As shown in Table \ref{table:adv-time}, the average time to fine-tune all  CodeBERT weights was approximately 8 hours. In contrast, fine-tuning CodeBERT with AdvFusion took approximately 5.5 hours, representing a reduction of about 44\% in training time compared to the full fine-tuning of the entire model.

We perform method name prediction on our baseline CodeLMs. The results are shown in Table~\ref{table:mnp}.
For this task, we observe that both AdapterFusion and AdvFusion have a larger impact on the results when they are added to GraphCodeBERT. This improvement is significant for all languages. 
For both models, AdvFusion slightly improves the results of AdapterFusion or achieves the same scores. We hypothesize that the variation could stem from the initial disparity in inputs and training methods between CodeBERT and GraphCodeBERT. GraphCodeBERT, utilizing dataflow graphs as input, gains a deeper understanding of the internal connections within code elements. This enhanced comprehension of the relationships among the programming languages enables GraphCodeBERT to suggest more effective method names by leveraging the knowledge from other programming languages for the language of the target task when AdvFusion is used.




\vspace{10pt}
\begin{mdframed} [backgroundcolor=gray!20, linewidth=1pt]
\textbf{When \textit{AdvFusion} is used for fine-tuning Code-LMs, we can achieve better or on par results compared to other PEFT methods. The improvement is observed more for the three programming languages Ruby, JavaScript, and Go for code summarization. As AdvFusion is a PEFT method, the training time is reduced and approximately 80\% fewer parameters are trained compared to full fine-tuning Code-LMs.}
\end{mdframed}

\subsection{RQ2: Languages' Contribution for a target Programming Language}

We assess the contribution of each language adapter across all programming languages for code summarization, comparing AdvFusion with AdapterFusion. Due to space constraints, we present only the results for Ruby, as the behaviour of other languages follows a similar trend. Figures illustrating the contributions of the other languages are available in the supplementary materials.

We extract the attention at AdvFusion and AdapterFusion when we fine-tune  CodeLMs+AdvFusion and CodeLMs+AdapterFusion, respectively (separate experiments). 
Fig.~\ref{fig:fusion-ruby} demonstrates the contribution of each language at each layer in CodeBERT+AdapterFusion when the Ruby test dataset is fed to the fine-tuned model. 
As discussed previously, it is noted that in most layers, a high percentage of attention (more than 80\%) is towards Ruby (the gray bar), rather than attending to other languages. In other words, not much is learned from other programming languages.

\vspace{3mm}

\begin{mdframed} [backgroundcolor=gray!20, linewidth=1pt]
\textbf{Programming languages could benefit from the other resourceful languages differently in different layers. Higher-resource languages do not necessarily contribute more to the low-resource language, such as Ruby.}
\end{mdframed}

\begin{figure}

\includegraphics[width=\columnwidth]{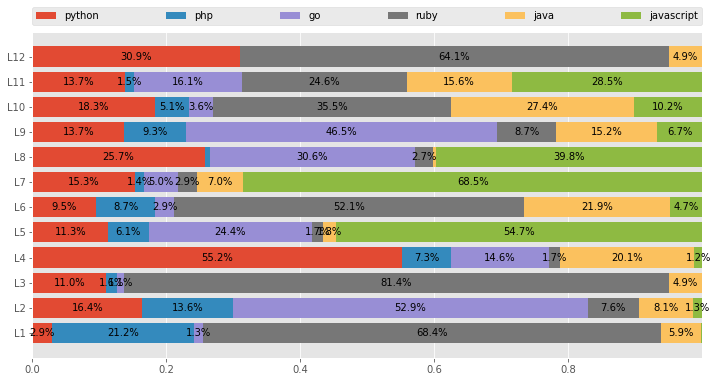}
\centering
\caption{The attention contribution from each programming language at each layer when we feed the Ruby test dataset to the fine-tuned AdvFusion model.}
\label{fig:adv-fusion-ruby}
\end{figure}

Fig.~\ref{fig:adv-fusion-ruby} shows the contribution of each language in CodeBERT + AdvFusion when the Ruby test dataset is fed to the fine-tuned model. The y-axis is the layer number in CodeBERT, and the x-axis shows the percentage of contribution of each language. Here, AdvFusion pays more attention to other programming languages. 
For instance, Ruby has the following learning: it learns more from Go in the second layer (i.e., $52.9\%$ of attention is grabbed from the Go adapter), it learns more from Python than Ruby in the fourth layer (i.e., $56.2\%$), and it learns more from JavaScript in layer seven. Even in the higher layers, learning from other languages is continued and the attention is distributed to other languages, and not only focused on Ruby. 
More interestingly, PHP is the most resourceful language in the dataset, but its contribution to Ruby is less than other languages. This suggests that there is no relationship between the size of the language dataset and its contribution to Ruby.


%% file: Sections/Discussion.tex
\section{Discussion} \label{section:discussion}

\textbf{\textit{When should we use adapters for monolingual fine-tuning? } }

In our experiments, we found that adapter-based fine-tuning is as effective as standard fine-tuning for high-resource languages in code summarization but more computationally efficient. It also enhances results for low-resource languages. Low-resource languages are the ones that have less training data available. Hence, we recommend adapter-based fine-tuning for monolingual fine-tuning in code summarization.
This result is similar to the findings in the literature \cite{liu2023empirical,weyssow2023exploring}.
We also observe that for Code-LMs in our study, adapters perform better than LORA and are a better choice among these two PEFT approaches.

However, for method name prediction on languages with limited resources, employing task adapters can still yield benefits without significant performance decline, while also reducing memory and time in fine-tuning.

\textbf{\textit{When should we consider knowledge transfer in multilingual fine-tuning?}}

Multilingual fine-tuning, as shown by \cite{ahmed2021multilingual}, often outperforms monolingual fine-tuning across resource levels. Table \ref{table:different_methods} highlights that some languages, like PHP, benefit less from multilingual adapters (e.g., AdapterFusion, AdvFusion) compared to full fine-tuning, possibly due to limited cross-language utility or insufficient PEFT parameter capacity. Python and Java show mixed results with PEFT, while AdvFusion effectively improves performance for low-resource languages by leveraging insights from Ruby and others.


\textbf{\textit{Which languages could a low-resource language take advantage of in a multilingual setting?}}

We observed that when using AdvFusion, Ruby has benefited from Go, Python and JavaScript, as depicted in Fig. \ref{fig:adv-fusion-ruby}. This study does not focus on the syntactic or semantic similarities between the source and target programming languages but rather on which languages are most useful for Ruby from the perspective of a model in practice. Continuation of other programming languages is provided in supplementary materials.

Fig.~\ref{fig:adv-sample} represents a heatmap generated from a Ruby sample fed into CodeBERT + AdvFusion. 
The x-axis displays Ruby tokens, while the y-axis shows the six programming languages of the CodeSearchNet dataset. Lighter colours indicate higher attention. This heatmap illustrates the attention each token receives from each programming language in the dataset.



The highest attention on the tokens is from other language adapters than the Ruby adapter; as observed, the attention from the Ruby adapter is very low (note the Ruby adapter row, which is dark everywhere).
However, for instance, the function signature of the sample, \code{sum},\code{(},\code{a},\code{b},\code{)} received more attention from Go rather than Ruby, and also the document tokens corresponded to the function signature, \code{the}, \code{sum}, and \code{of}, are paid more attention by Go. 
This is aligned with our observations in Fig. \ref{fig:adv-fusion-ruby}, as discussed in RQ2.

\textbf{\textit{When can adapters be helpful, in terms of architectures and tasks?}}

In our study, incorporating adapters into the CodeT5 baseline for code summarization led to a performance decline. We attribute this to the pre-existing decoder stack in CodeT5, which may limit adaptability compared to models like CodeBERT or GraphCodeBERT that train the decoder from scratch. A similar issue was reported in \cite{wang2023oneAdapter}, where CodeT5 fine-tuning underperformed relative to CodeBERT and GraphCodeBERT.

\textbf{\textit{Which target tasks could benefit from multilingual fine-tuning using AdvFusion? }}\\
We have conducted experiments on code summarization and method name prediction, demonstrating the effectiveness of AdvFusion. We hypothesize that other tasks with consistent output modalities across datasets could similarly benefit from AdvFusion and AdapterFusion architectures. For instance, tasks like code review and commit message generation—where the output is natural language—could leverage multilingual fine-tuning, provided there are datasets available in multiple programming languages.

\begin{figure}

\includegraphics[width=\columnwidth]{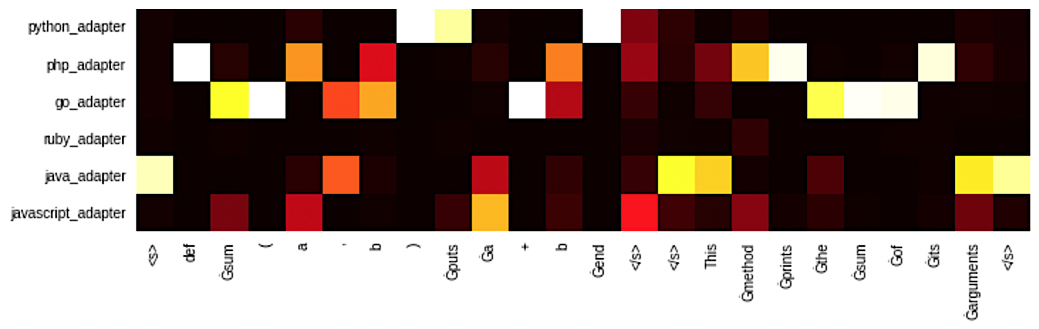}
\centering
\caption{AdvFusion's attention heatmap across six language adapters for a Ruby sample in the fine-tuned model. X-axis displays code tokens, while the y-axis shows attention from each adapter.}
\label{fig:adv-sample}
\end{figure}

%% file: Sections/RelatedWork.tex
\section{Related Work}
\label{section:related-work}


\subsection{Parameter Efficient Fine Tuning Studies}

PEFT methods offer an alternative to fully fine-tuning language models and have been widely applied in NLP tasks\cite{houlsby2019parameter,pfeiffer2020mad,pfeiffer2020adapterfusion,hu2021lora}. Adapter-based fine-tuning is often shown to outperform full fine-tuning, particularly in zero-shot, cross-lingual, and low-resource scenarios, by the authors of \cite{low-res-related}. Meanwhile, several  experiments have been done on various PEFT methods, such as Adapters, LoRA, and Prefix-Tuning, and evaluated their performance, scalability, and knowledge transfer across more than 100 NLP tasks by \cite{related-ding2023parameter}.




Research on PEFT approaches in software engineering is extensive\cite{goel2022cross,weyssow2023exploring,tse2023,liu2024mftcoder}. An empirical study on natural language to code transferability using adapters was conducted by the authors of \cite{goel2022cross}. PEFT methods such as LoRA \cite{hu2021lora} and prompt tuning in code generation were explored in \cite{weyssow2023exploring}, with a focus on their advantages in large language models compared to small ones. Prompt tuning's impact on CodeBERT and CodeT5 on code tasks such as defect prediction, summarization, and translation were investigated by \cite{tse2023}. They compared fully fine-tuned and prompt-tuned models, assessing accuracy and data efficiency. Other work proposed a multi-task fine-tuning framework using PEFT methods \cite{liu2024mftcoder}. The performance of PEFT approaches on Just-In-Time Defect Prediction (JIT-DP) and Commit Message Generation (CMG) is evaluated in \cite{liu2024delving}.


\subsection{Language Models}

In the past few years, there is a lot of effort on representing source code using deep learning models for different applications such as code generation \cite{zeng2022extensive,zhou2022doccoder,fried2022incoder}, code summarization \cite{gu2022assemble,ahmed2022learning,nie2022impact}, program synthesis \cite{vaithilingam2022expectation,nijkamp2022conversational,ellis2021dreamcoder,austin2021program}, code search \cite{nadeem2022codedsi}, and bug repair \cite{2022arXiv220700301R,2022arXiv220805446Z}.
A number of models are also released that are pre-trained on source code and/or code and comment with different objective functions, which are then fine-tuned on multiple downstream tasks \cite{wang2021codet5,feng2020codebert,guo2020graphcodebert} such as code summarization \cite{feng2020codebert,wang2021codet5,lu2021codexglue,ahmed2021multilingual}. Examples of these models include CodeT5 \cite{wang2021codet5}, CodeT5+ \cite{wang2023codet5+}, PLBART \cite{ahmad2021unified}, and CodeGPT. Each has versions fine-tuned for specific downstream tasks.


Recent code-focused LLMs include Deepseek-Coder2\cite{zhu2024deepseek}, CodeLlama \cite{code-llama-roziere2023code}, an extension of Llama2 \cite{touvron2023llama}, and AlphaCode \cite{alphacode-li2022competition}, a competition-level code generator. The BigCode community developed StarCoder and StarCoderBase \cite{li2023starcoder}, models with 15.5 billion parameters and advanced capabilities like infilling and efficient large-batch inference. Though these large code-LLMs have billions of parameters and differ from models like CodeBERT, Authors of \cite{ma2024unveiling} suggest code-LMs perform better than code-LLMs in identifying syntax, making fine-tuning code-LLMs more challenging. In paper \cite{dou2023towards}, it is shown that code-LMs often outperform code-LLMs in tasks like clone code detection.

\textbf{Differences:} Research on code language models (code-LMs) primarily focuses on program understanding and generation across tasks \cite{zeng2022extensive}, with limited attention to their adaptability across different languages for specific software engineering tasks. Recent studies explore code-LM transferability for Ruby code summarization \cite{chen2022transferability}, using few-shot learning, code-related prompts, or project-specific data \cite{2022arXiv220704237A, 2022arXiv220601335B, 2022arXiv220407501K, 2022arXiv220600804A}. However, none aim to leverage knowledge from multiple languages for a target language. This paper empirically analyzes monolingual and multilingual adapters for two tasks and proposes an architecture to harness cross-language insights.

%% file: Sections/Threats.tex
\section{Threats to Validity}
\label{section:threat}


\textbf{External Validity:} 
This study examines code summarization and method name prediction, focusing on the languages in the CodeSearchNet dataset, which may limit generalizability. We tested our method on CodeT5p, CodeBERT and GraphCodeBERT but performance may vary with other encoder-decoder or decoder-only models. Additionally, the representativeness of CodeSearchNet code samples may not fully capture real-world programming scenarios, affecting the applicability of our findings. \textbf{Internal Validity:} 
 Fine-tuning pre-trained models requires careful hyperparameter selection, which can be challenging. To address this, we adopted adapter-specific hyperparameters from \cite{pfeiffer2020adapterfusion}, minimizing tuning risks. We also ensured consistency by using identical dataset splits, evaluation metrics, and trainable parameter budgets across all models, reducing variability and enabling reliable comparisons.
 \textbf{Construct Validity:} We evaluate our code summarization model with BLEU-4 score and method name prediction with F1-Score metrics. Choosing metrics is crucial, different ones yield different results. We have selected these metrics based on current literature, minimizing evaluation threats.
{Please note that, although there is research about the weaknesses of BLEU score for code summarization \cite{roy2021reassessing,haque2022semantic}, it is intuitive and easy to understand, as it measures the overlap between the generated translation and one or more reference translations based on n-grams. Secondly, BLEU is language-independent, making it applicable across various language pairs without the need for language-specific tuning. Also, as the results are all reported in the same metric, it is a fair comparison and will not threaten the validity of the results.} \textbf{Conclusion Validity:}
 We validated our study's conclusions by rigorously testing research queries RQ1 and RQ2 with different random seeds and two backbone models. Focusing on code summarization and method name prediction, this process assures the reliability of our findings.

%% file: Sections/Conclusion.tex
\section{Conclusion and Future Works}
\label{section:conclusion}

There is a recent interest in the software engineering research community to use parameter-efficient fine-tuning for code-LMs. 
In this work, we first evaluate and report the performance of AdapterFusion compared to multilingual full fine-tuning. 
We then introduced AdvFusion, a new architecture that promotes knowledge transfer from other programming languages in a multilingual fine-tuning setting, to enhance the learning capability of AdapterFusion. Using AdvFusion, the performance is on par with or is improved while involving a smaller parameter budget. The training time is also reduced.


Future directions of this work are applying AdvFusion on other downstream tasks and code-LMs with different architectures. 
Another avenue could be investigating the reasons that one programming language has a higher attention score to the target task language, compared to other languages. This latter study could open avenues for training models that benefit from other languages for a target task in a low-resource language.